\documentclass[aps,prb,twocolumn,showpacs]{revtex4}
\usepackage{graphicx}

\begin{document}

\title{Nature of the inhomogeneous state of the extended t-J model on a square lattice}

\author{Chung-Pin Chou and Ting-Kuo Lee}
\affiliation{Institute of Physics, Academia Sinica, Nankang, Taipei
11529, Taiwan}

\begin{abstract}
We carry out the variational Monte Carlo calculation to examine
spatially inhomogeneous states in hole- and electron-doped cuprates.
By using Gutzwiller approximation, we consider the excitations,
arising from charge density, spin density and pair field, of the
mean-field ground state of the $t-J$ model. It leads to the stripe
patterns we have found numerically in a generalized $t-J-$type model
including mass renormalization from the electron-phonon coupling. In
the hole-doped side, a robust $d$-wave superconducting order results
in the formation of the half-doped antiferromagnetic
resonating-valence-bond (AF-RVB) stripes shown by the well-known
Yamada plot. On the other hand, due to a long-range AF order in
electron-doped materials, a stripe structure with the "in-phase"
magnetic domain (IPMD) is obtained in the underdoped regime instead
of the AF-RVB stripe. The IPMD stripe with the largest period
permitted by lattice size is stabilized near the underdoped region
and it excludes the Yamada plot from electron-doped cases. Based on
finite lattice size to which we can reach, the existence of IPMD
stripes may imply an electronic phase separation into an
electron-rich and an insulating half-filled AF long-range ordered
domains in electron-doped compounds.
\end{abstract}

 \maketitle


Since the discovery of high-temperature superconductivity in the
layered cuprate materials, there have been many evidences for stripe
structures in several families of the hole-doped cuprates, e.g.,
$La_{2-\delta}Sr_{\delta}CuO_{4}$
\cite{TranquadaNat95,KivelsonRMP03}. One of the many puzzles of the
stripes is the doping dependence of the incommensurate magnetic
peaks associated with the stripes measured by neutron-scattering
experiments which obeys the so-called Yamada plot
\cite{YamadaPRB98}. It represents the existence of the half-doped
stripe with average of $1/2$ hole in one charge modulation period at
$1/8$ hole density and below. There have been many early theoretical
works attempting to explain the Yamada plot
\cite{MartinsPRL00,WhitePRL98,ArrigoniPRB02}. One possible scenario
for such correlation is that tendency for charges toward phase
separation can lead to various structures including stripes, puddles
\cite{ZaanenPRB89,EmeryPhysC93}, or even cluster glasses with
randomly-oriented stripe domains recently observed by scanning
tunneling spectroscopy (STS) \cite{KohsakaSci07,ParkerNat10}.
Recently we have used a variational Monte Carlo (VMC) technique
\cite{CPCPRB10} to successfully establish the half-doped stripes in
the extended $t-J-$type Hamiltonian by including a
mass-renormalization effect due to a weak electron-phonon coupling.

So far the experimental situation in electron-doped materials is
much less clear \cite{ArmitageRMP10}. There are several indirect
evidences for a homogeneous state in electron-doped compounds coming
from measurements of neutron scattering and core-level photoemission
spectra \cite{HarimaPRB01,MotoyamaPRL06}. Yamada \textit{et al.}
have reported only commensurate spin fluctuations observed by
neutron scattering in the superconducting (SC)
$Nd_{1.85}Ce_{0.15}CuO_{4}$ \cite{YamadaPRL03}. It is different from
the incommensurate peaks observed in hole-doped cuprates which are
considered as the hall mark of the "out-of-phase" stripe domains
with a $\pi$-phase-shifted staggered magnetic moment. Instead of
stripes, the short-range spatial inhomogeneity of the
antiferromagnetic (AF) correlations was recently reported for  the
electron-doped superconductor $Pr_{0.88}LaCe_{0.12}CuO_{4-\delta}$
by Zhao \textit{et al.} \cite{ZhaoNatPhys11} by using STS and
neutron scattering. In addition, there are also evidences to support
inhomogeneous states from measurements of muon spin rotation
($\mu$SR) \cite{SonierPRL03,KlaussJPCM04}, nuclear magnetic
resonance \cite{ZamborszkyPRL04}, magnetoresistance
\cite{FournierPRB04}, and thermal conductivity \cite{SunPRL04}.
Whether Inhomogeneity in electron-doped compounds is intrinsic or
induced by the cerium doping or oxygen defects were discussed by two
recent works \cite{DaiPRB05,HigginsPRB06}. All these results suggest
that the possibility of phase separation and inhomogeneity is an
unresolved issue in electron-doped cuprates.

On the theoretical side, there are some numerical evidences that
Hubbard models produce stripes in the electron-doped system. Within
an unrestricted Hartree-Fock approach, the occurrence of the
diagonal filled stripes having average of one doped electron per
stripe site has been demonstrated earlier \cite{SadoriPRL00}. Later,
the vertical "in-phase" stripe domains without $\pi$-phase-shifted
staggered magnetic moment in the $t-t'$ Hubbard model have been
found in the electron-doped regime \cite{ValenzuelaPRB06}. The
unusual doping evolution of the Fermi surface detected by
angle-resolved photoemission spectroscopy \cite{ArmitagePRL02} was
explained by assuming the inhomogeneous in-phase stripe phases
\cite{GranathPRB04} but without including strong correlations. Since
the strong correlation makes the difference in energies between
uniform states and various hole-doped stripe states very small which
has been shown recently \cite{CPCPRB08,CPCPRB10}, it is beyond the
accuracy of Hartree-Fock method to address this kind of difference.
Therefore, a careful variational approach is needed to examine the
stability of the electron-doped stripe states.

In this paper, we shall first demonstrate that the particular
spatial patterns of modulations of the charge density, spin density
and pair field could be derived by considering the excitations of
the mean-field ground state of the extended $t-J$ model using
Gutzwiller approximation. Once the relations between charge, spin
and pair field are revealed we then explicitly construct the stripe
wave functions. As a comparison with what we have done previously
for the hole-doped cases \cite{CPCPRB10}, we shall add an
electron-phonon interaction to the model before carrying out the
numerical calculations. Just as before, we will not consider the
full effect of electron-phonon coupling but only examine the
simplest effect of mass renormalization of charges due to phonon
couplings. The renormalization effect depending on the local carrier
density is treated self-consistently in the VMC method taking into
account the strong correlation exactly \cite{CPCPRB10}. We find that
half-doped stripes obtained in the hole-doped systems are no longer
stable in the electron-doped cases for a range of electron-phonon
interaction strength. Instead the system is very likely to have an
electronic phase separation. The different results between
hole-doped and electron-doped systems is mainly due to the strong AF
long-range order in the latter system. The effect of $t'/t$ will be
also discussed.


We consider the extended $t-J$ Hamiltonian on a square lattice given
by
\begin{eqnarray}
H=-\sum_{i,j,\sigma}t_{ij}\left(\tilde{c}_{i\sigma}^{\dag}\tilde{c}_{j\sigma}+H.c.\right)+J\sum_{\langle
i,j\rangle}\mathbf{S}_{i}\cdot\mathbf{S}_{j}, \label{e:equ1}
\end{eqnarray}
where the hopping $t_{ij}=t$, $t'$, and $t''$ for sites i and j
being the nearest, second-nearest, and third-nearest neighbors,
respectively. Other notations are standard. In the following, the
bare parameters $t''$ and $J$ in the Hamiltonian are set to be
$(t'',J)/t=(-t'/2,0.3)$. Since doubly-occupied sites in
electron-doped materials play the same role as holes in hole-doped
cases, we treat the hole- and electron-doped cases in the same
manner except that $t'/t\rightarrow-t'/t$ and
$t''/t\rightarrow-t''/t$ \cite{LeePRL03}. $t'/t<0$ ($>0$)
corresponds to the hole-doped (electron-doped) regions. In this
paper, we primarily study the hole- and electron-doped phase
diagrams for different $t'/t$.

In a generalized mean-field theory to include the possibility of
spatially non-uniform solutions, we define the local carrier density
$\rho_{i}$, the local AF order $m_{i}$, and the nearest-neighbor
pair field $\Delta_{ij}$. The mean-field Hamiltonian is simply given
by
\begin{eqnarray}
\hat{H}_{MF}=\left(
     \begin{array}{cc}
       c_{i\uparrow}^{\dag} & c_{i\downarrow} \\
     \end{array}
   \right)\left(
            \begin{array}{cc}
              H_{ij\uparrow} & D_{ij} \\
              D_{ji}^{*} & -H_{ji\downarrow} \\
            \end{array}
          \right)\left(
                   \begin{array}{c}
                     c_{j\uparrow} \\
                     c_{j\downarrow}^{\dag} \\
                   \end{array}
                 \right),\label{e:equ2}
\end{eqnarray}
where the matrix elements
\begin{eqnarray}
H_{ij\sigma}&=&-\delta_{j,i+\hat{1}}-t'_{v}\delta_{j,i+\hat{2}}-t''_{v}\delta_{j,i+\hat{3}}\nonumber\\
&+&\rho_{i}+\sigma m_{i}(-1)^{x_{i}+y_{i}}-\mu_{v},\nonumber\\
D_{ij}&=&\Delta_{ij}\delta_{j,i+\hat{1}}.\label{e:equ3}
\end{eqnarray}
Here $\hat{1}$, $\hat{2}$, and $\hat{3}$ correspond to the nearest,
second-nearest, and third-nearest neighbors, respectively, and
$\sigma=\uparrow(1)$ or $\downarrow(-1)$.

Once the variational parameters $\rho_{i}$,
$m_{i}$, and $\Delta_{ij}$ are given, we can diagonalize
Eq.(\ref{e:equ2}) to obtain $N$ positive and $N$ negative
eigenvalues with corresponding eigenvectors $(u_{i}^{n},v_{i}^{n})$
and $(\bar{u}_{i}^{n},\bar{v}_{i}^{n})$, given by
\begin{eqnarray}
\left(
  \begin{array}{c}
    \gamma_{n} \\
    \bar{\gamma}_{n} \\
  \end{array}
\right)=\left(
          \begin{array}{cc}
            u_{i}^{n} & v_{i}^{n} \\
            \bar{u}_{i}^{n} & \bar{v}_{i}^{n} \\
          \end{array}
        \right)\left(
                 \begin{array}{c}
                   c_{i\uparrow} \\
                   c_{i\downarrow}^{\dag} \\
                 \end{array}
               \right).\label{e:equ5}
\end{eqnarray}
Here $N$ is the lattice size. We can formulate the trial wave
function fixing the number of electrons $N_{e}$ with the Gutzwiller
projector $P_{G}$ and the hole-hole repulsive Jastrow factor $P_{J}$
(see the details of Ref.~\onlinecite{CPCPRB08}),
\begin{eqnarray}
|\Psi\rangle&=&P_{G}P_{J}P_{N_{e}}\prod_{n}\gamma_{n}\bar{\gamma}_{n}^{\dag}|0\rangle\nonumber\\
&\propto&P_{G}P_{J}P_{N_{e}}\prod_{n}\sum_{i}\left(u_{i}^{n}f_{i}^{\dag}+v_{i}^{n}d_{i}^{\dag}\right)|\tilde{0}\rangle.\label{e:equ6}
\end{eqnarray}
To avoid the divergent determinant because of the presence of nodes
in the RVB-type wave functions with periodic boundary condition, a
particle-hole transformation \cite{YokoyamaJPSJ88,CP12},
$c_{i\uparrow}^{\dag}\rightarrow f_{i}$ and
$c_{i\downarrow}^{\dag}\rightarrow d_{i}^{\dag}$, has been
introduced in Eq.(\ref{e:equ6}). Here
$|\tilde{0}\rangle\equiv\prod_{i}f_{i}|0\rangle$.

In principle, $\rho_{i}$, $m_{i}$, and $\Delta_{ij}$ could be
determined variationally. In practice, it is very difficult to
optimize the energy of such multi-variable problems. If we postulate
that the inhomogeneous states are actually fluctuations beyond the
uniform mean-field solution, then a more efficient method to find
the most probable solutions is to use Gutzwiller approximation
\cite{FukushimaPRB08}. In this approximation, the total energy
$\langle H\rangle$ with respect to the Gutzwiller-projected wave
function can be written as
\begin{eqnarray}
&&-\sum_{<i,j>,\sigma}t_{ij}g_{t}^{\sigma}(i)g_{t}^{\sigma}(j)\left(\chi_{ij}^{\sigma}+H.c.\right)\nonumber\\
&&-J\sum_{<i,j>}g_{s}(i)g_{s}(j)\left(\frac{3}{8}\left(\chi_{ij}\chi_{ij}^{\ast}+\Delta_{ij}\Delta_{ij}^{\ast}\right)-m_{i}m_{j}\right),\nonumber\\
\label{e:equ100}
\end{eqnarray}
where the Gutzwiller factors $g_{t}^{\sigma}(j)$ and $g_{s}(i)$ are
known to be \cite{WHKoPRB07}
\begin{eqnarray}
g_{t}^{\sigma}(i)&=&\sqrt{\frac{n_{i}(1-n_{i})(1-n_{i\bar{\sigma}})}{(1-n_{i\sigma})(n_{i}-2n_{i\uparrow}n_{i\downarrow})}},\nonumber\\
g_{s}(i)&=&\frac{n_{i}}{n_{i}-2n_{i\uparrow}n_{i\downarrow}}.\label{e:equ101}
\end{eqnarray}
Here $n_{i}=\sum_{\sigma}n_{i\sigma}=1-\delta_{i}$ and
$n_{i\sigma}=\frac{1-\delta_{i}}{2}+\sigma m_{i}$.
$\chi_{ij}(=\sum_{\sigma}\chi_{ij}^{\sigma}=\sum_{\sigma}\langle
c_{i\sigma}^{\dag}c_{j\sigma}\rangle_{0})$, $m_{i}(=\langle
S_{i}^{z}\rangle_{0})$, and $\Delta_{ij}(=\langle
c_{i\downarrow}c_{j\uparrow}-c_{i\uparrow}c_{j\downarrow}\rangle_{0})$
is the bond order parameter, staggered magnetization, and pair field
with respect to the non-projected wave function, respectively. For a
usual mean-field theory, these parameters are assumed to be constant
and the same for all sites or bonds. Here we shall go one step
further by examining the fluctuations beyond the constant mean-field
values as given by
\begin{eqnarray}
\delta_{i}&\rightarrow&\bar{\delta}+d\delta_{i}\nonumber\\
m_{i}&\rightarrow&\bar{m}+dm_{i}\nonumber\\
\Delta_{ij}&\rightarrow&\bar{\Delta}+d\Delta_{ij}.\label{e:equ102}
\end{eqnarray}
Here $d(...)$ means the small fluctuation away from the average
value $\overline{(...)}$. By substituting Eq.(\ref{e:equ102}) into
Eq.(\ref{e:equ100}), we obtain several coupled terms with the form
$d\delta_{i}dm_{i}^{2}$, $\bar{\Delta}d\delta_{i}d\Delta_{ij}$,
$\bar{m}d\delta_{i}dm_{i}$, and $d\delta_{i}d\Delta_{ij}^{2}$. There
are also non-coupled terms of the form $dm_{i}^{2}$,
$d\delta_{i}d\delta_{j}$, and $d\Delta_{ij}^{2}$. We neglect the
fluctuation of bond order and also skip the derivation of all the
terms as they are irrelevant for our calculations below.

In the hole-doped side, at finite doping there is no long-range AF
order ($\bar{m}=0$) but with a nonzero $d$-wave SC order parameter
($\bar{\Delta}\neq0$). Thus, according to the discussion above, the
relevant fluctuation terms to couple the charge, spin and pair
fields are $d\delta_{i}dm_{i}^{2}$ and
$\bar{\Delta}d\delta_{i}d\Delta_{ij}$. The Fourier transform of
these two terms are $d\delta_{q}dm_{-q/2}^{2}$ and
$\bar{\Delta}d\delta_{q}d\Delta_{-q}$, respectively. Hence the
modulation period of charge density, $a_{c}$, should equal to the
period of pair field, $a_{p}$, but is only half the period of the
spin density, $a_{s}$, {\it i.e.}, $a_{c}=a_{p}=a_{s}/2$. More
precisely the wave functions to include these fluctuations can have
order parameters of the form:
\begin{eqnarray}
\rho_{i}&=&\rho_{v}\cos\left[\frac{2\pi}{a_{c}}\left(y_{i}-\frac{1}{2}\right)\right],\nonumber\\
m_{i}&=&m_{v}^{M}\sin\left[\frac{2\pi}{a_{s}}\left(y_{i}-\frac{1}{2}\right)\right],\nonumber\\
\Delta_{i,i+\hat{x}}&=&\Delta_{v}^{M}\cos\left[\frac{2\pi}{a_{p}}\left(y_{i}-\frac{1}{2}\right)\right]-\Delta_{v}^{C},\nonumber\\
\Delta_{i,i+\hat{y}}&=&-\Delta_{v}^{M}\cos\left[\frac{2\pi}{a_{p}}y_{i}\right]+\Delta_{v}^{C}.\label{e:equ4}
\end{eqnarray}
Here $a_{c}=a_{p}=a_{s}/2$. The variational parameters are indicated
by subscript "$v$". The average charge density is determined by the
chemical potential not included in $\rho_{i}$. This is exactly the
wave function called the AF resonating-valence-bond (AF-RVB) stripe
state by us in Ref.~\onlinecite{CPCPRB08}. There we had shown that
the variational energy of the uniform $d$-wave RVB ($d$-RVB) state
($\rho_{v}=m_{v}^{M}=\Delta_{v}^{M}=0$) is considered as the
reference energy. Furthermore if we add a weak electron-phonon
interaction to the extended $t-J$ model, as shown in
Ref.~\onlinecite{CPCPRB10}, the half-doped AF-RVB stripes have lower
energy than the $d$-RVB state. This AF-RVB stripe pattern is in good
agreement with experiments as well
\cite{TranquadaNat95,AbbamonteNatPhys05}.

The electron-phonon interaction is introduced by assuming the
hopping terms $t_{ij}$ in Eq.(\ref{e:equ1}) modified due to the
spatial variation in carrier density in the sense that the sites
with larger carrier density in the modulated charge pattern have
larger $t_{ij}$ \cite{CPCPRB10}. Then we assume a linear relation
between the electron-phonon coupling strength $\lambda$ and doping
density $\delta$, $\lambda(\delta)/\lambda(0)=1-3\delta\equiv
f(\delta)$. Now $t_{ij}$ can be renormalized to
\begin{eqnarray}
t_{ij}^{\ast}=t_{ij}\left(1-\frac{\Lambda}{2}\left[f(n_{\delta}^{i})+f(n_{\delta}^{j})\right]\right),\label{e:equ0}
\end{eqnarray}
where $n_{\delta}^{i}$ is the carrier density at site $i$ and
$\Lambda$ the bare parameter in the Hamiltonian. The details are
given in Ref.~\onlinecite{CPCPRB10}. In what follows, we will use a
notation $"t^{\ast}-J"$ to stand for this Hamiltonian.

In the electron-doped side, it had been shown \cite{ShihCJP07} that
there is a robust long-range AF order ($\bar{m}\neq0$) in the
underdoped regime in contrast to hole-doped systems. Based on our
previous discussion of fluctuations using the Gutzwiller
approximation, the important contributions in Eq.(\ref{e:equ100})
will be $\bar{m}d\delta_{i}dm_{i}$ and
$\bar{\Delta}d\delta_{i}d\Delta_{ij}$. Hence, the magnetic
modulation has the same period as charge and pair field, that is,
$a_{c}=a_{p}=a_{s}$. This pattern is very different from AF-RVB
stripe states discussed above for hole-doped systems. The wave
function to describe such a new pattern is called the in-phase
magnetic domains (IPMD). The IPMD stripe state has the same function
of $\rho_{i}$ and $\Delta_{ij}$ as Eq.(\ref{e:equ4}), only the
magnetic modulation is now changed to
\begin{eqnarray}
m_{i}=-m_{v}^{M}\cos\left[\frac{2\pi}{a_{s}}\left(y_{i}-\frac{1}{2}\right)\right]+m_{v}^{C},\label{e:equ7}
\end{eqnarray}
where a finite staggered magnetic moment $m_{v}^{C}$ is included.
Due to the presence of long-range AF order for electron-doped
systems, we have four variational states to be considered. These are
the pure $d$-RVB uniform SC state, the AF-RVB stripe state, the IPMD
stripe state and the uniform state with the coexistence of
long-range AF and SC orders (AF+SC) \cite{ShihCJP07}. The AF+SC
state can be simply constructed by setting
$\rho_{v}=m_{v}^{M}=\Delta_{v}^{M}=0$ in Eqs.(\ref{e:equ4}) and
(\ref{e:equ7}).


\begin{figure}[top]
\rotatebox{0}{\includegraphics[height=2.5in,width=3.5in]{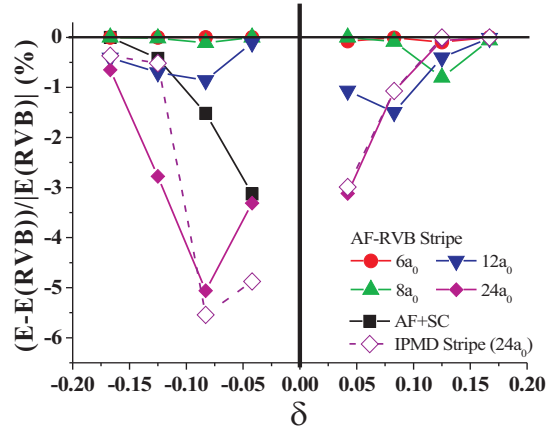}}
\caption{(Color online) The doping dependence of the optimized
energy difference in terms of percentage for several trial states
denoted in the figure. The reference ground state is the uniform
$d$-RVB state. The positive (negative) doping density $\delta$ means
the hole-doped (electron-doped) cases. Note that the AF-RVB stripe
states with different magnetic periods are indicated. The lattice
size $N$ is $24\times24$. Here $\Lambda=0.25$ and
$|t'/t|=0.1$.}\label{Fig1}
\end{figure}

Figure \ref{Fig1} shows the percentage of energy change with respect
to the $d$-RVB state for three other trial wave functions in the
extended $t^{\ast}-J$ model with smaller $|t'/t|(=0.1)$ for hole-
and electron-doped phase diagrams. In the hole-doped case, the
half-doped AF-RVB stripe is still observed like our previous studies
for $t'/t(=-0.2)$ \cite{CPCPRB10}. In fact, the half-doped AF-RVB
stripe can be always found for all $|t'/t|$ we have investigated in
the hole-doped region as shown in Fig.\ref{Fig3}(a) and (b). In
other words, the Fermi surface topology seems not to affect the
stability of the half-doped stripe obtained in the extended
$t^{\ast}-J$ model.

\begin{figure}[top]
\rotatebox{0}{\includegraphics[height=4in,width=3.5in]{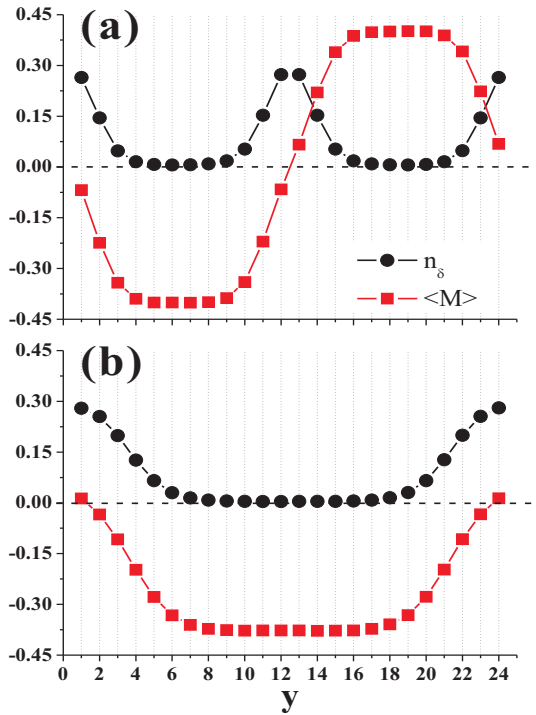}}
\caption{(Color online) The profile of carrier density $n_{\delta}$
and staggered magnetization $\langle M\rangle$ for the optimized
states: (a) the AF-RVB stripe and (b) the IPMD stripe with $24a_{0}$
magnetic periodicity at $1/12$ electron-doping in the extended
$t^{\ast}-J$ Hamiltonian. The bare parameters $\Lambda=0.25$ and
$|t'/t|=0.1$. All quantities are calculated in a $24\times24$
lattice.}\label{Fig2}
\end{figure}

In the left panel of Fig.\ref{Fig1} we find the IPMD stripe states
are stabilized for electron-doping less than $0.1$. For doping
greater than $0.1$, the AF-RVB stripe state with the largest
magnetic period has the lowest energy ($24a_{0}$ is the largest size
we have studied here). The magnetic period does not change with the
doping density as the half-doped stripes. We also examine the IPMD
stripe states with different periods in addition to $24a_{0}$. The
results are not shown here, but the most stable IPMD stripe has the
largest magnetic period as lattice size. The difference between the
AF-RVB and IPMD stripes for the maximum magnetic period of $24a_{0}$
is actually negligible. Since the IPMD state has much lower energy
for doping less than $0.1$, it appears that the $t^{\ast}-J$
Hamiltonian does not prefer to break the system into many
$\pi$-phase-shift magnetic domains. In Fig.\ref{Fig2}(a) and (b) for
$1/12$ electron density, the spatial variation of charge density and
staggered magnetization along the direction of modulation are shown
for the AF-RVB stripe and IPMD stripe, respectively. For both
states, there is almost no doped carriers in the region with the
strongest staggered magnetization. It is essentially a phase
separated state with an electron-rich region and an insulating
half-filled AF long-range ordered region. The IPMD stripe state has
staggered magnetization $|\langle M\rangle|=0.265$, which is close
to the homogeneous AF+SC state ($|\langle M\rangle|=0.269$). This is
consistent with the commensurate short-range spin correlations
detected by neutron scattering \cite{YamadaPRL03}.

\begin{figure}[top]
\rotatebox{0}{\includegraphics[height=4in,width=3.5in]{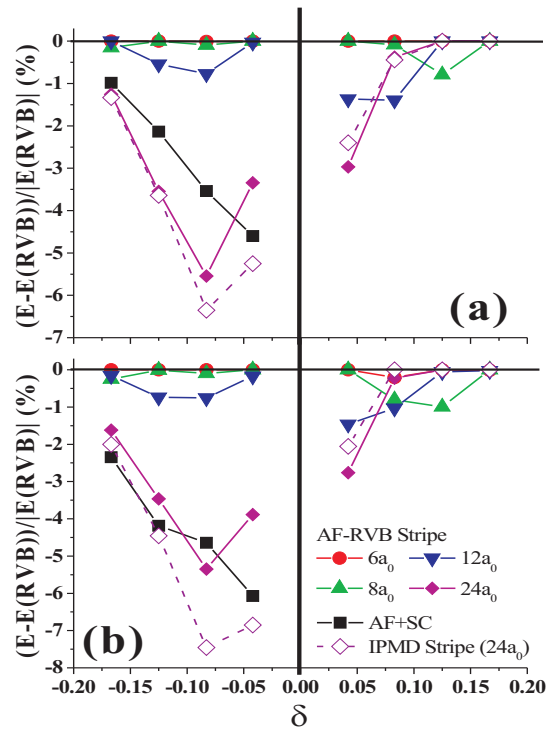}}
\caption{(Color online) The same descriptions are presented in the
caption of Fig.\ref{Fig1} except for (a) $|t'/t|=0.2$ and (b)
$|t'/t|=0.3$.}\label{Fig3}
\end{figure}

For larger $t'/t$, we expect robust AF correlations should persist
to the higher doping in electron-doped cases. Figure \ref{Fig3}
shows this common feature for the uniform AF+SC and the IPMD stripe
states. In Fig.\ref{Fig3}(a), while the homogeneous AF+SC state has
much lower energy than the $d$-RVB state, the IPMD stripe state is
still the best candidate for electron-doping $\delta<0.1$. It is
worth pointing out that for electron-doping $\delta>0.1$ the IPMD
stripe state begins to gain energy to compete with the AF-RVB stripe
state in the case of $t'/t=0.2$. Interestingly, as increasing
$t'/t(=0.3)$ further, the AF-RVB stripe states entirely disappear in
the electron-doped phase diagram possibly due to its
$\pi$-phase-shift domains in the spin modulation, as shown in
Fig.\ref{Fig3}(b). Except for the higher electron-doping region
($\delta>0.15$) where the homogeneous AF+SC state still exists, the
IPMD stripe state is dominant within the wider doping range
($0<\delta<0.15$). In addition to these stripe states shown in the
phase diagram, we have also examined other trial wave functions with
different shapes and periods for stripes, such as glassy stripes
with randomly-oriented $8\times8$ magnetic patches, IPMD stripes
with other periods, and the diagonal stripe with the largest
magnetic period $(12\sqrt{2}a_{0})$. However, none of them can be
stabilized for all $t'/t$ in electron-doped cases (not shown).

\begin{figure}[top]
\rotatebox{0}{\includegraphics[height=2.5in,width=3.5in]{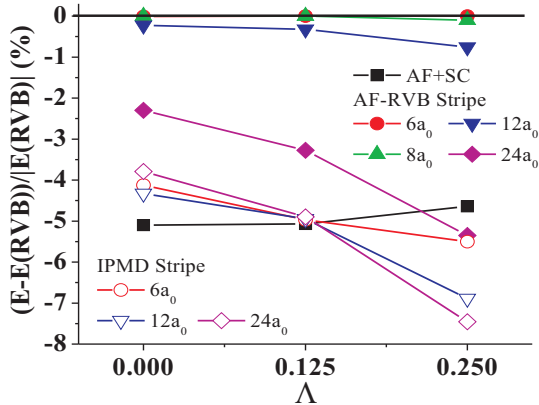}}
\caption{(Color online) $\Lambda$-dependence of the optimized energy
difference in terms of percentage for several trial states denoted
in the figure at 1/12 doping. Here $N=24\times24$ and
$t'/t=0.3$.}\label{Fig4}
\end{figure}

To complete the discussion on the VMC results, we study the effect
of the electron-phonon coupling strength $\Lambda$ on the phase
diagrams. Since the IPMD stripe state has much lower energy near
$1/12$ electron-doping, we only show $\Lambda$-dependence of the
optimized energy difference for several trial wave functions
including IPMD stripes with different magnetic periods at $1/12$
doping in Fig.\ref{Fig4}. As we expected, the stability of the
uniform AF+SC state is almost independent of $\Lambda$ as
$\Lambda\lesssim0.2$. If the electron-phonon coupling strength is
smaller $(\Lambda<0.125)$, the AF+SC state would be still a good
candidate for the ground state near the underdoped region in the
electron-doped side. However, once $\Lambda$ becomes larger, IPMD
stripe states begin to be stabilized. Notably, the best one for
$\Lambda=0.25$ is either the IPMD stripe state with the largest
magnetic period or the phase-separated state. Unlike hole-doped
cases, there is no place for AF-RVB stripe states at this doping in
electron-doped phase diagrams.

Although the lowest energy solutions of electron-doped systems for
the extended $t^{\ast}-J$ Hamiltonian are IPMD stripes and not
AF-RVB stripes as the hole-doped systems, the particular patterns of
modulation periods of charge density, spin density and pair field
for both stripes are derived from the fluctuations of the order
parameters used in the mean field theory as shown by using the
Gutzwiller approximation. The inhomogeneous states like stripes or
phase separation is really due to excitations or fluctuations of
order parameters of the ground states of the strongly correlated
$t-J$ model. When electron-phonon interaction is introduced to
renormalize the mass of carriers, these inhomogeneous solutions then
become stable.


In conclusion, following the same approach we have used to study
stripes for hole-doped systems \cite{CPCPRB08,CPCPRB10}, we have
further studied the possibilities of having non-uniform ground
states for the electron-doped cases using the VMC method. By using
the Gutzwiller approximation to examine the fluctuations of order
parameters used to obtain the $d$-RVB ground states, we have found
that the period of modulation of charge density, staggered moment
and pair field are correlated. For hole-doped systems the lowest
energy stripes, the AF-RVB stripe states, have the period of
staggered moment twice of that of charge and pair field. These
stripes with half a carrier per charge domain become the ground
states after we include the weak electron-phonon coupling. However,
similar approach for electron-doped cases has turned out a different
stripe pattern. The lowest energy stripes, the IPMD stripe states,
now have the period of staggered moment same as that of charge and
pair field. One of the main reasons is that for a large electron
doping range the ground state has long-range AF order. The numerical
results show that at low doping, the IPMD stripe state with the
largest magnetic period same as lattice size is composed of an
electron-rich region and an insulating half-filled long-range AF
ordered region. We believe that it may indicate a phase separated
state in electron-doped compounds. This statement requires the
numerical calculation with large lattice size that is left for our
future study.

Our result that the electron-doped systems should not have the
half-doped stripes observed in hole-doped systems is consistent with
the neutron scattering results for cuprates that Yamada plot is only
observed for hole-doped cuprates \cite{YamadaPRB98} but not for
electron-doped \cite{ZhaoNatPhys11}. The evidence on magnetic
inhomogeneity we have found numerically has been indirectly observed
in the electron-doped cuprates $Pr_{2-\delta}Ce_{\delta}CuO_{4}$ and
$Pr_{0.88}LaCe_{0.12}CuO_{4-\delta}$ by using $\mu$SR and STS
experiments, respectively \cite{KlaussJPCM04,ZhaoNatPhys11}. The
fact that for electron doping we predict possible phase separation
which is also a hotly debated issue by experiments on cuprates is
quite interesting. We look forward to possible resolution of this
issue in the near future.


\begin{acknowledgments}
This work was supported by the National Science Council in Taiwan
with Grant No. 98-2112-M-001-017-MY3. The calculations are performed
in the National Center for High-performance Computing and the PC
Cluster III of Academia Sinica Computing Center in Taiwan.
\end{acknowledgments}

\end{document}